%% file: prd.tex
\newif\ifprd
\prdtrue 

\ifprd
\documentclass[aps,prd,twocolumn,superscriptaddress,showpacs,nofootinbib]{revtex4}
\else
\documentclass[12pt,twoside,letterpaper,doublespace]{article}
\fi

\usepackage{graphicx}
\usepackage{dcolumn}
\usepackage{bm}
\usepackage{lineno}
\usepackage{setspace}
\usepackage{amssymb}
\usepackage{xspace}

\newcommand{\ppbar}{\ensuremath{p\bar{p}}\xspace}
\newcommand{\zzero}{\ensuremath{Z^{0}}\xspace}
\newcommand{\zzeros}{\ensuremath{Z^{0}}s\xspace}
\newcommand{\zzerozzero}{\ensuremath{Z^{0}Z^{0}}\xspace}
\newcommand{\ifb}{\ensuremath{\mathrm{fb}^{-1}}\xspace}

\newcommand{\pt}{\ensuremath{p_{\mathrm{T}}}\xspace}
\newcommand{\Et}{\ensuremath{E_{\mathrm{T}}}\xspace}

\newcommand{\hadem}{\ensuremath{E_{HAD}/ E_{EM}}\xspace}

\newcommand{\tev}{\ensuremath{\mathrm{\,Te\kern -0.1em V}}\xspace}
\newcommand{\gev}{\ensuremath{\mathrm{\,Ge\kern -0.1em V}}\xspace}
\newcommand{\mev}{\ensuremath{\mathrm{\,Me\kern -0.1em V}}\xspace}
\newcommand{\kev}{\ensuremath{\mathrm{\,ke\kern -0.1em V}}\xspace}
\newcommand{\ev}{\ensuremath{\mathrm{\,e\kern -0.1em V}}\xspace}
\newcommand{\gevc}{\ensuremath{{\mathrm{\,Ge\kern -0.1em V\!/}c}}\xspace}
\newcommand{\mevc}{\ensuremath{{\mathrm{\,Me\kern -0.1em V\!/}c}}\xspace}
\newcommand{\gevcc}{\ensuremath{{\mathrm{\,Ge\kern -0.1em V\!/}c^2}}\xspace}
\newcommand{\mevcc}{\ensuremath{{\mathrm{\,Me\kern -0.1em V\!/}c^2}}\xspace}
\newcommand{\roots}{\ensuremath{\sqrt{s}}\xspace}

\newcommand{\cdfii}{CDF~II\xspace}
\newcommand{\cdf}{CDF\xspace}
\newcommand{\abseta}{\ensuremath{|\eta|}\xspace}
\newcommand{\mee}{\ensuremath{m_{ee}}\xspace}
\newcommand{\meeee}{\ensuremath{m_{eeee}}\xspace}
\newcommand{\chisq}{\ensuremath{\chi^2}\xspace}
\newcommand{\kovermp}{\ensuremath{k/\overline{M}_{Pl}}\xspace}

\def\pythia      {\mbox{\sc pythia}\xspace}
\def\geant      {\mbox{\sc geant}\xspace}
\def\herwig      {\mbox{\sc herwig}\xspace}

\newcommand{\gtozzeeee}{\ensuremath{G\rightarrow{Z^{0}Z^{0}}~\rightarrow~eeee}\xspace}

\begin{document}

\ifprd
\else
\linenumbers
\doublespace
\fi

\ifprd
\else
\hfill CDF/PUB/EXOTIC/CDFR/8806

\hfill Version 3.3
\vspace*{1.5cm}
\fi

\ifprd
\input{July_2007_Authors_Visitors}
\title{Search for New Heavy Particles decaying to $\zzerozzero \to eeee$ in \ppbar Collisions at \roots = 1.96 \tev}
\else
\begin{center}
\begin{Large}
Search for New Heavy Particles Decaying to
$\zzerozzero \to eeee$ in \ppbar
Collisions at \roots = 1.96 \tev
\end{Large}
\end{center}
\vspace{1in}
\begin{center}
\begin{large}{The \cdf Collaboration\\}
\end{large}
\today
\end{center}
\fi

\ifprd
\vspace*{2.0cm}
\fi
\begin{abstract}
We report the results of a search for the anomalous production of a massive
particle decaying to four electrons via two \zzero bosons in 1.1~\ifb of
\ppbar collisions at \roots=1.96 \tev collected by the \cdfii detector at
Fermilab.  We employ optimized electron identification criteria to maximize
acceptance and efficiency.  We estimate the backgrounds in the invariant
mass range 500 - 1000~\gevcc to be $0.028 \pm 0.009$ (stat)
$\pm~0.011$~(syst) events.  We observe zero events in this search region.
Assuming a Randall-Sundrum graviton production model, we set 95\% CL limits
on $\sigma \times \mathrm{BF}(G\rightarrow \zzero \zzero) < 4 - 6$ pb,
depending on the graviton mass.

\end{abstract}

\pacs{12.60.Cn,13.85.Rm,14.70.Hp}

\ifprd
\maketitle
\else
\newpage
\fi

\section{\label{sec:Intro}Introduction}

We present a search for new heavy particles ``$G$'' in the decay mode
\gtozzeeee in $\bar{p}p$ collisions at $\sqrt{s}=1.96$ \tev performed with
the \cdfii detector at the Fermilab Tevatron.  Previous analyses of double
gauge boson production at the Tevatron have focused on standard model
production
\cite{Abulencia:2007tu,Abazov:2005ys,Affolder:2001cd,Acosta:2005pq}.
Here, we present for the first time a search for massive particles $G$ which
decay to \zzerozzero which could indicate physics beyond the standard model.

The goal of this search is to be sensitive to production of any massive
particle which could decay to \zzerozzero, that is, to avoid focusing on any one specific model;
however, for the purpose of quantifying acceptance for this search,
we consider the virtual production of gravitons in a simple
Randall-Sundrum (RS1) scenario~\cite{Randall:1999ee,Randall:1999vf}.  In this model,
the geometry consists of two 3-branes which confine the standard model
sector separated from each other by a single extra dimension.
One can look for evidence of the extra dimension at particle colliders in
the form of a Kaluza-Klein tower of discrete, massive gravitons.
In the RS1 scenario, the gravitons predominantly decay to
jets, and the remaining modes are $W^+W^-$ ($\sim 10$\%), \zzerozzero ($\sim
5$\%), $\gamma\gamma$ ($\sim 5$\%), and $ll$ ($\sim 2$\% per
lepton)\cite{Allanach:2002gn}.
Searches for the decays of such particles to photons and electrons have
been performed~\cite{:2007kpa,:2007sb,Abazov:2005pi}.
If the couplings to
leptons and photons are suppressed relative to the couplings to gauge
bosons~\cite{Fitzpatrick:2007qr}, such a particle might escape detection in
these searches.  Here, we have searched for massive particles in
their decays to \zzero bosons.

In the leptonic final states of \zzero decay, the expected signal from the
model described above is small, as are the backgrounds.  In order to
maximize acceptance and efficiency for the four-electron signature, we use
optimized calorimetric electron identification criteria, select electron
candidates identified as isolated tracks where there is no calorimeter
coverage, and use low electron energy thresholds.

The organization of this article is as follows: first, we describe the
components of the \cdfii detector relevant to this search and summarize the
data sample and event selection criteria.  Then we describe the 
background estimation, report the results of the search, and interpret the
results in the context of the lightest massive RS1 graviton.

\section{The \cdfii Detector\label{sec:cdf}}
This analysis uses 1.1~\ifb of \ppbar collisions collected by the \cdfii
detector, a general purpose magnetic spectrometer.  We briefly describe the
components of the detector relevant to this search here.  A complete
description can be found elsewhere~\cite{Acosta:2004yw}.

A combination of tracking systems reconstruct the trajectories and measure
momenta of charged particles in a 1.4T solenoidal magnetic field.
Trajectories of charged particles are reconstructed using an eight-layer
silicon microstrip vertex tracker~\cite{Sill:2000zz} at radii $1.3 < r < 29$
cm~\cite{coordinates}, and a 96-layer open-cell drift chamber (COT)
providing eight superlayers of alternating axial and stereo position
measurements~\cite{Affolder:2003ep}.  The COT allows track reconstruction at
large radii $43 < r < 132$~cm in the region $\abseta < 1.6$, and provides
full geometric coverage for $\abseta < 1.0$.

Outside the tracking volume, segmented electromagnetic (EM)
lead-scintillator and hadronic (HAD) iron-scintillator sampling calorimeters
measure particle energies~\cite{Balka:1987ty}.  In the central region
($\abseta < 1.1$), the calorimeters are arranged in a projective-tower
cylindrical geometry, divided azimuthally into $15^{\circ}$ wedges which
measure EM energies with a resolution of $[\sigma(E)/E]^2 =
(13.5\%)^2/E_T + (2\%)^2$.  In the region covered by the forward calorimeter
($1.1 < \abseta < 3.6$), the calorimeters are arranged in an
azimuthally-symmetric disk geometry and measure EM energies with a
resolution of $[\sigma(E)/E]^2 = (16.0\%)^2/E + (1\%)^2$.  Wire chambers
(scintillator strips) embedded in the central (forward) EM calorimeters at
the electromagnetic shower maximum ($\sim 6 X_0$) provide position and
lateral shower development measurements used to identify electrons by their
characteristic energy-deposition distribution.  The beam luminosity is
determined by measuring the inelastic
\ppbar collision rate with gas Cherenkov detectors\cite{clc}, located in the
region $3.7 < \abseta < 4.7$. 

\section{\label{sec:Data}Event Selection}
Events are selected for collection by a three-level trigger system.  We
search in data collected by triggering on a central high-momentum electron.
Each of two trigger paths used requires an energy cluster in the central
calorimeter and a track which projects to the energy cluster.  The primary
trigger path requires a clustered transverse energy $\Et>18$~\gev,
transverse momentum of the associated track $\pt>9$~\gevc, the ratio of
energy measured in the hadronic to electromagnetic calorimeters,
$\hadem<0.125$ and a lateral shower profile consistent with an electron.
The second trigger path requires a cluster with $\Et > 70$~\gev and an
associated track with $\pt>15$~\gevc.

We select events containing one ``seed'' electron which satisfies trigger
requirements and those listed in Table \ref{tab:triggers}, and three other
electrons which satisfy either the central, forward, or track requirements
in Table \ref{tab:triggers}.  To maximize acceptance and efficiency for
events containing four electrons, we select three other electron candidates
using optimized identification and kinematic criteria in the central or
forward calorimeters, and from isolated tracks pointing to uninstrumented
regions of the calorimeters.

Electron candidates are formed in the central and forward calorimeters
from isolated energy deposits with $\Et \geq 5$~\gev.  An electron is
considered to be isolated in the calorimeter if the sum of the transverse
energy detected within a cone $\Delta R\equiv\sqrt{(\Delta\eta)^2 +
(\Delta\phi)^2} \le 0.4$, minus the electron $\Et$, is less than 20\% of the
electron $\Et$.  For clusters in the central calorimeter, where tracking
efficiency is optimal, we require that a track reconstructed in the COT
project to the cluster.  Tracks must include measurements in at least three
axial and two stereo superlayers of the COT, and the track coordinate along
the beam direction, $z_0$, must also lie within the nominal extent of the
interaction region, $|z_0| < 60$ cm.  To reduce backgrounds from hadrons
misidentified as electrons, central candidates must also satisfy an
energy-dependent requirement $\hadem<0.055+0.00045 \gev^{-1}
\times E_{EM}$. Forward candidates must have $\hadem < 0.05$ and lie within
$1.1 < \abseta < 2.5$.

The ``seed'' electron candidate must satisfy the above requirements to be
reconstructed in the central calorimeter, and satisfy additional
selection criteria imposed by the trigger.  Specifically, the ``seed'' electron
must also have $\Et \geq 20$~\gev, satisfy the same lateral shower profile
requirement as the triggering one, and have an associated track with
$\pt \geq 10$~\gevc.

\begin{table*}
\caption[Electron Selection Criteria]
{Electron Selection Criteria}
\label{tab:triggers}
\begin{center}
\begin{tabular}{cccc}
\hline
Criteria &  Central (Seed)  & Forward & Track \\
\hline
\hline
\Et (\gev)		& $\geq 5(20)$                 & $\geq 5$ &  \\
$|z_0|$ (cm) 	& $<60 $ 			&   & $< 60 $ \\
\hadem       		& $<0.055+0.00045 \gev^{-1} \times E$	&  $< 0.05$ &  \\
Isolation   		& $<0.2 $                       & $ < 0.2  $ & $> 0.9$ \\
$p_T$ (\gevc) 	& $ (\geq 10)$ 			&  & $\geq 10$\\
\hline
\end{tabular}
\end{center}
\end{table*}

Some calorimeter acceptance for electrons is lost in 24 one-degree gaps in
$\phi$ between the central calorimeter projective wedges, a region at $0.7 <
\eta < 1.0$ and $75^{\circ} < \phi < 90^{\circ}$ which accommodates cables
and cryogenic utilities for the solenoid, a gap between central calorimeter
arches at $\eta = 0$, and the gap at the junction between the central and
forward calorimeters at $1.0 < \abseta < 1.2$.  Together, these regions add
up to approximately 17\% of the solid angle for $\abseta <1.2$, which for a
four-lepton final state represents an acceptance loss of approximately half.

We recover acceptance by forming electron candidates from isolated tracks
which project to the gaps between instrumented regions of the calorimeter.
A track is defined to be isolated in the tracking chamber if the transverse
momentum of the track is more than 90\% of the total transverse momentum of
all tracks within a cone $\Delta R
\le 0.4$ around the candidate track.  We require track electron
candidates be consistent with originating from prompt decays by requiring
that they pass within 0.2mm (2mm) of the axial beam position for tracks with
(without) position measurements in the silicon vertex tracker.

To reconstruct \zzero candidates, we form all unique combinations of pairs of
electrons in the event.  To avoid rejecting events where the charge of one
electron is misidentified, we do not impose an opposite charge requirement on
the pair.  We ensure that all electron candidates are isolated from each
other by requiring a separation of 0.2 in $\Delta R$ between any two
electron candidates in the combination.  If both candidates in a pair have
associated tracks, we ensure they are consistent with originating from the
same parent by requiring their $z_0$ measurements to lie within 5 cm of each
other.  The invariant mass distributions for events containing \zzero
candidates formed from a ``seed'' electron candidate together with just one
other electron candidate and the subset where an isolated track is used as
the second electron candidate are shown in Fig.~\ref{fig:ee_yield}.

\begin{figure}[tb]
\centering
\mbox{\includegraphics[width=7.3cm,clip=]{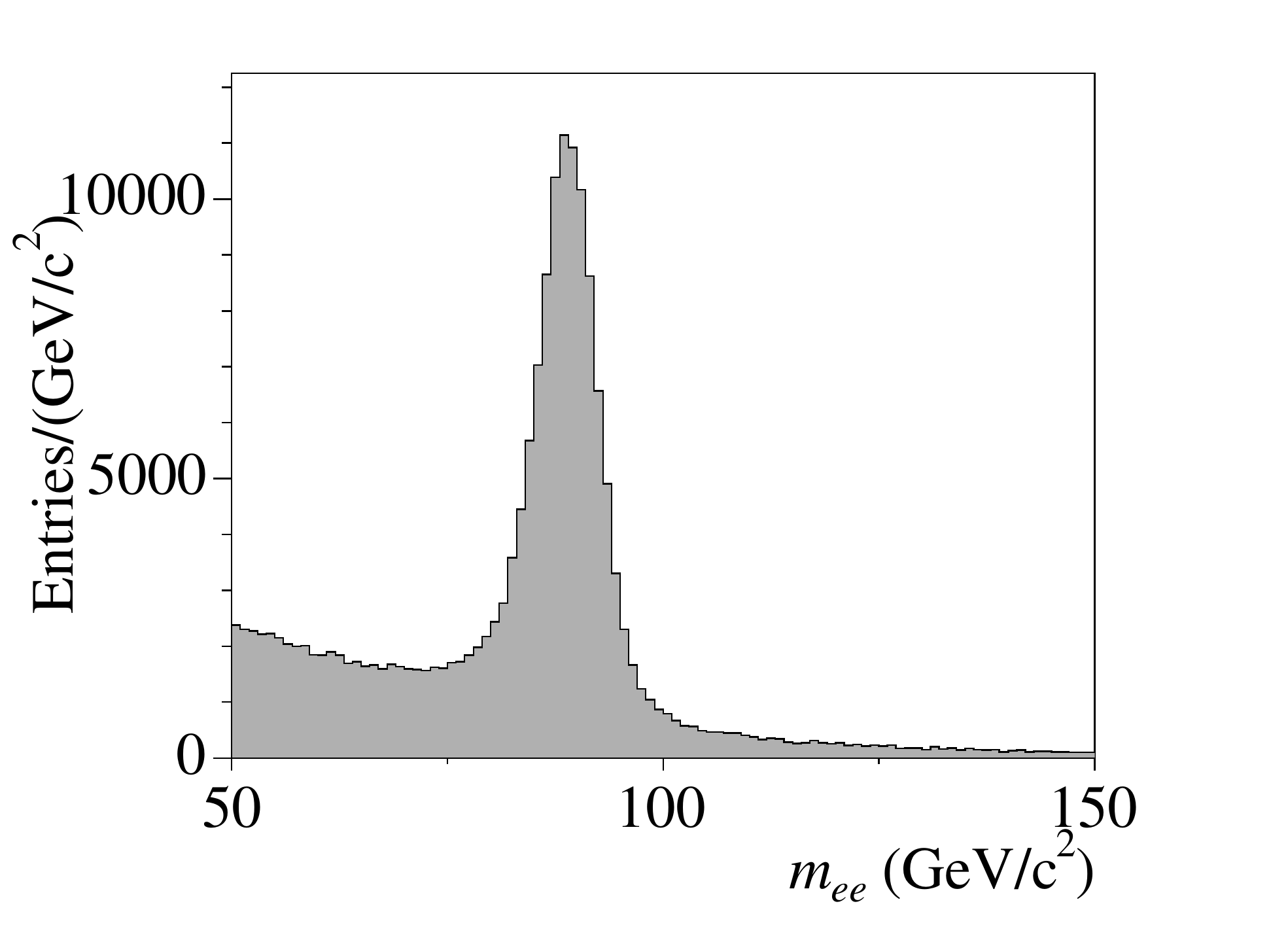}
\put(-50,120){(a)}
}
\mbox{\includegraphics[width=7.3cm,clip=]{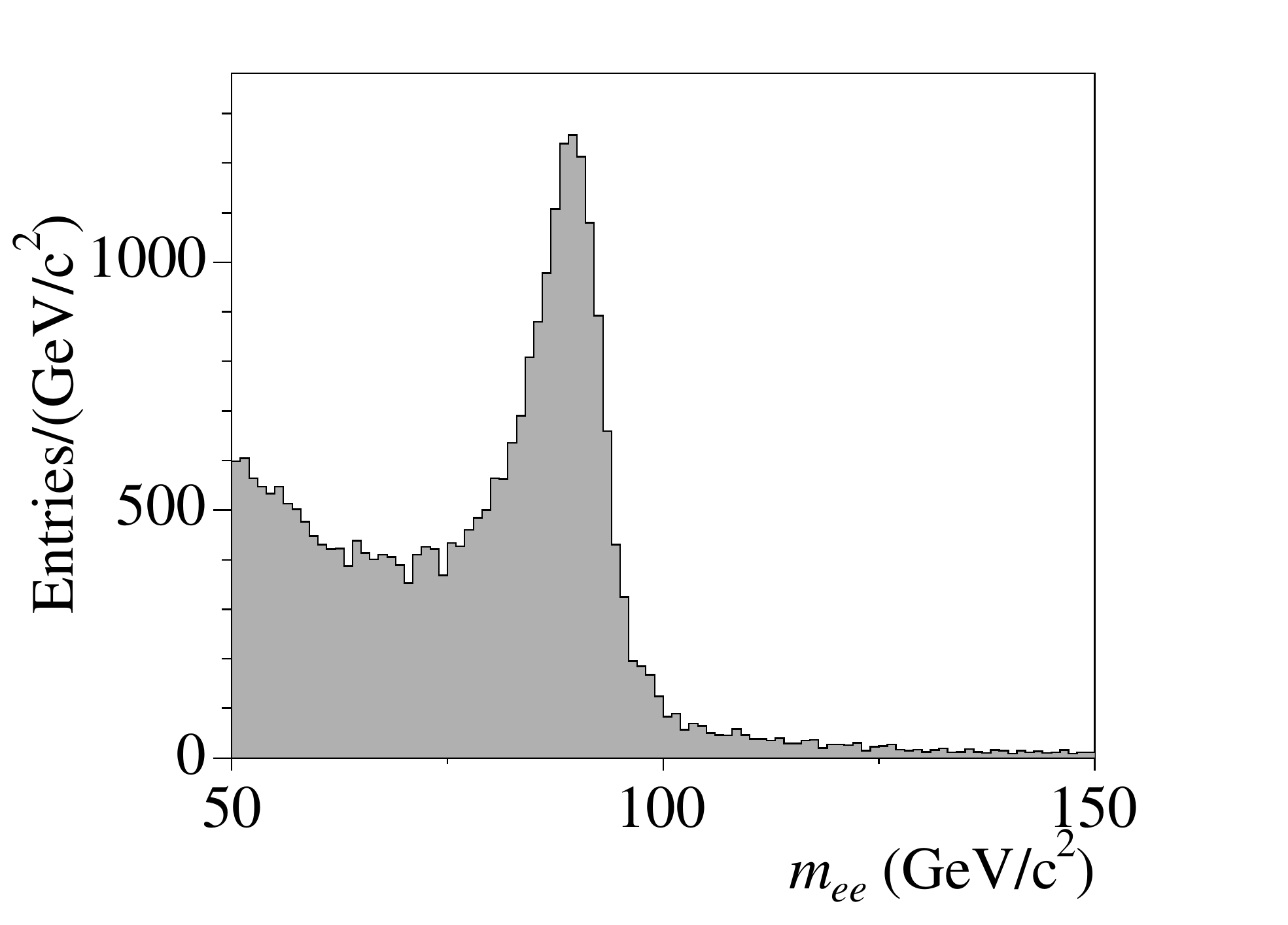}
\put(-50,120){(b)}
}
\caption{Distribution of \mee for \zzero candidates
formed from a ``seed'' electron candidate together with a second electron
candidate (a), and the subset of \zzero candidates formed from a ``seed''
electron candidate and an isolated track (b).\label{fig:ee_yield}}
\end{figure}

To reconstruct \zzero pairs, we form all unique combinations of all \zzero
candidates containing a ``seed'' electron with all other \zzero candidates
in the event and again require a separation of 0.2 in $\Delta R$ between any
two electrons in the four-electron combination.

The variable
\begin{linenomath}
\begin{eqnarray}
\chisq & = & 
\sum_{i=1,2}\left(\frac{m_i - m_{\zzero}} {\sigma_{i}} \right)^2
\end{eqnarray}
\end{linenomath}
quantifies consistency between a given combination and a $\zzerozzero \to
eeee$ final state, where $m_\mathrm{\zzero} = 91.19$~\gevcc is the nominal
\zzero mass~\cite{Yao:2006px}, $m_i$ is the measured invariant mass of each
candidate \zzero in the combination computed from the electron candidates'
four-momenta, and $\sigma_{i} $ is the uncertainty on the mass of each
\zzero candidate consisting of a contribution from the intrinsic width of
the
\zzero boson and a contribution propagated from the individual electron
energy or momentum measurements (typically $\sim 3.5$~\gevcc).  In
each event, we retain the one \zzerozzero combination with the lowest \chisq.  

We fix the final event selection criteria before examining the event yield
in the signal region.  We define the signal region as the events containing
a four electron combination with $\chisq<50$ and $\meeee > 500$~\gevcc.
Events with $\meeee < 400$~\gevcc are used as a low-mass control region to
estimate backgrounds, as described in
Section~\ref{sec:background_estimation}.
We find 12 events containing four electron candidates in this control region.

Although we do not focus on one specific model, for the purpose of
quantifying acceptance for this signature, we consider a graviton-production
scenario implemented in the \herwig~\cite{Corcella:2000bw} Monte Carlo generator
which is treated in a model-independent way,
assuming only that there is a universal coupling of the graviton
to standard-model particles.
For comparison, we interpret the couplings in the context of the RS1 model.
We determine geometric and kinematic acceptance and reconstruction
efficiency for this model
using Monte Carlo calculations followed by
a \geant-based simulation of the CDF II detector~\cite{Brun:1987ma}.
We consider graviton production followed by
decay to two \zzero bosons followed by decay into four electrons.  We use a
leading-order calculation implemented in
\herwig to estimate acceptance
times efficiency for the model.  For a RS1 graviton with mass $M_{G} =
500$~\gevcc and ratio of warp factor to Planck mass,
\kovermp = 0.1, Fig.~\ref{fig:gzzee50_chisq} shows the distribution of
reconstructed
\chisq and \meeee, the invariant mass of the \zzerozzero combination with
the lowest \chisq computed from the four-momenta of the two
\zzero candidates.
\begin{figure}[tb]
\centering
\mbox{\includegraphics[width=7.3cm,clip=]{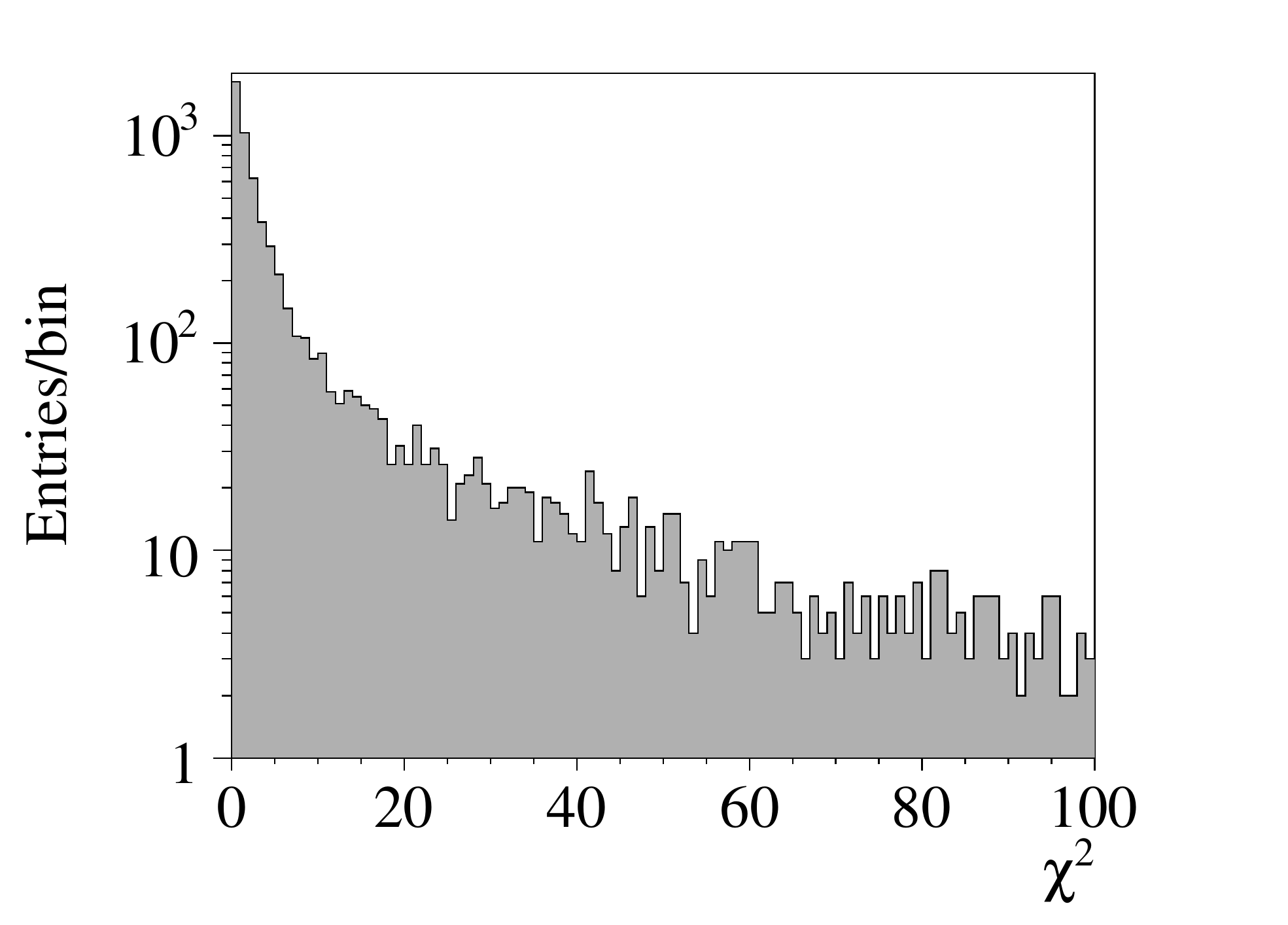}} 
\mbox{\includegraphics[width=7.3cm,clip=]{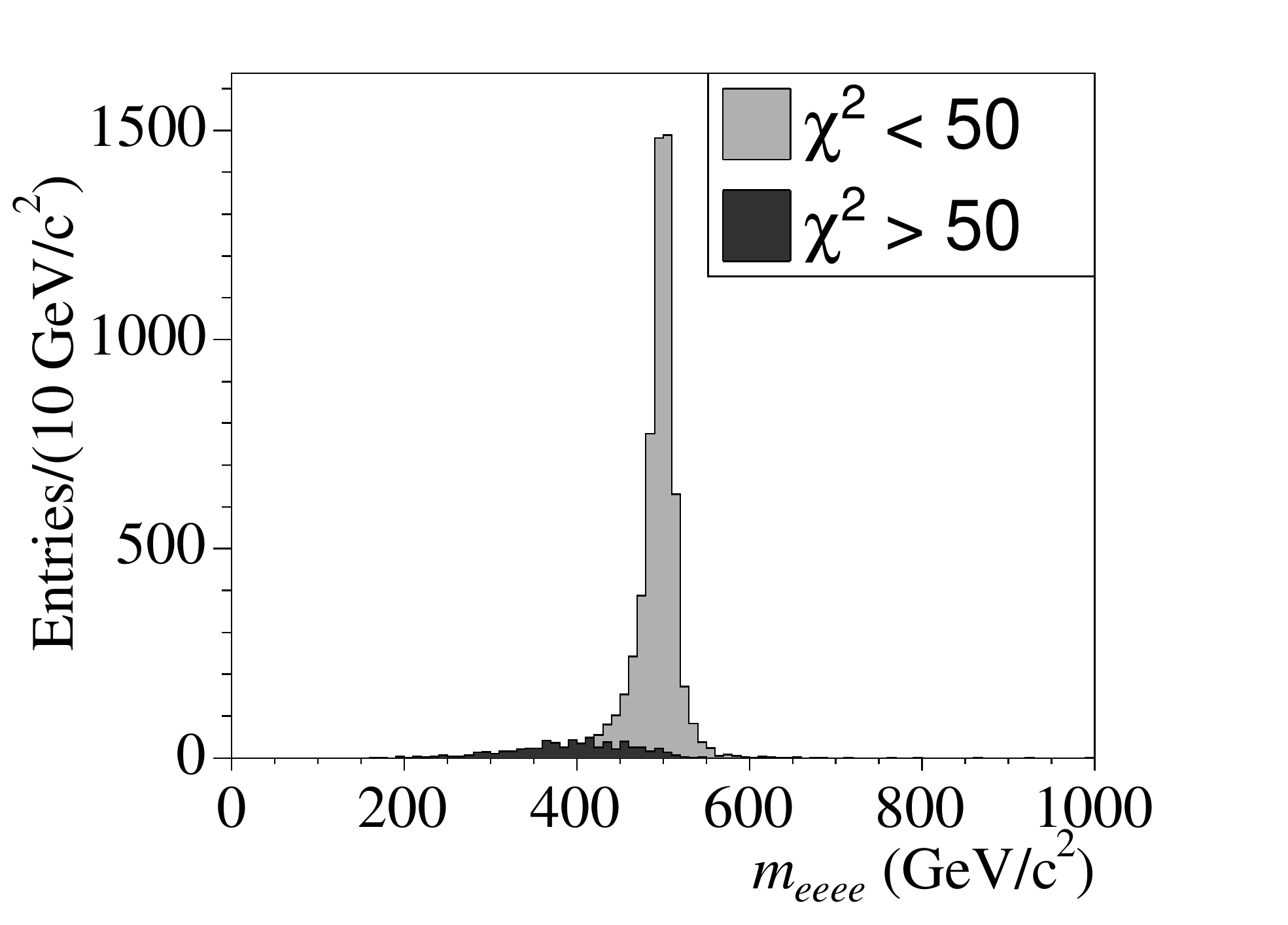}} 
\caption{Distribution of \chisq for simulated Randall-Sundrum signal scenario
($m_G=500~\gevcc$) (top).
Four-electron invariant mass distribution for
events satisfying $\chisq < 50$ (gray) and for events which fail this
requirement (black) (bottom).
\label{fig:gzzee50_chisq}}
\end{figure}
As expected for events containing two real \zzero
bosons,
the \chisq distribution peaks near zero, and
the total invariant mass of selected combinations is centered on the
generated graviton mass $(500~\gevcc)$.
Events which contain mis-measured electrons contribute to
the population with large \chisq values.
The width of the \meeee distribution, $\sim 15$ \gevcc,
is dominated by the detector resolutions of the constituent
electron candidates.
We find the geometric and kinematic acceptance times efficiency
for this model to be 65\%.
Of the events reconstructed,  93\% yield a four electron combination
with $\chisq < 50$.

The total acceptance versus $\meeee$ is shown in
Fig.~\ref{fig:acc}.  
At very high graviton mass, the momentum of the
daughter \zzeros becomes significant, which can cause the electrons to have
a small opening angle and fail the isolation requirement.

\begin{figure}[tb]
\centering
\mbox{\includegraphics[width=7.3cm,clip=]{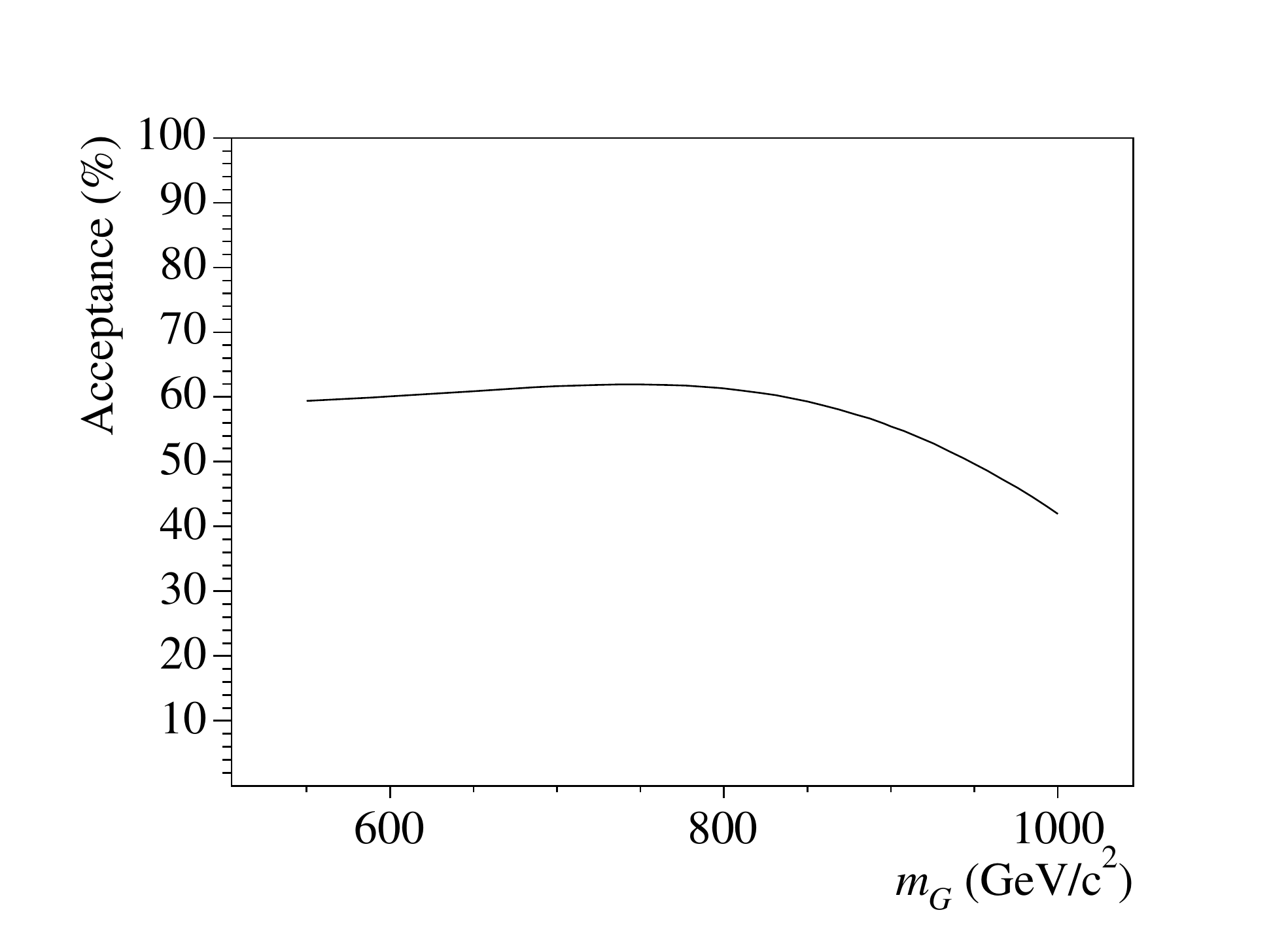}} 
  \caption{Acceptance for Randall-Sundrum graviton decaying to \zzerozzero
 versus its mass.\label{fig:acc}}
\end{figure}

\section{Background Estimation}\label{sec:background_estimation}

In studies using Monte Carlo simulation to estimate the main sources of
backgrounds, we find that the dominant background consists of events in
which one or more hadrons satisfy the electron ID requirements in the
four electron combination.  

We use control samples in the data to obtain the shape and
normalization of this background in the signal region.  Background-dominated
(hadron-enriched) control samples are selected from the data.
We form hadron candidates, $h$, from calorimeter clusters in a manner
identical to the central and forward electron candidates, with two
exceptions.  The hadron candidate must fail the relevant $\hadem$ criterion,
and to increase the size of the control samples, we do not impose any
isolation requirements.  In Fig.~\ref{fig:trigfake} we show the invariant
mass of all pairings of one seed electron candidate with one hadron
candidate.
The absence of a significant peak at the \zzero mass indicates that
contamination from electrons in the hadron candidates is small.\footnote{The
presence of a small amount of electron contamination in the hadron candidate
sample has a negligible effect on the estimate of the background at high
mass, and is included in the systematic uncertainty we assign to the
background estimation method.}

We obtain five control samples,
namely the four-electron  sample which has $\meeee < 400\gevcc$ introduced
above, and additional control samples having one, two, three, or four hadron
candidates by repeating the \zzerozzero selection
procedure, forming combinations using one or more hadron candidates with
electron candidates and retaining the minimum \chisq combination for each
event.  The distributions of the minimum \chisq versus \meeee for samples
with different numbers of hadron candidates are shown in
Fig.~\ref{fig:data_chisq_meeee}.
For reconstructed masses smaller than twice the \zzero mass, there is a
correlation between \chisq and mass caused by a kinematic threshold effect.
At higher masses, the two variables are much less correlated.

\begin{figure}[tb]
\centering
\mbox{\includegraphics[width=7.3cm,clip=]{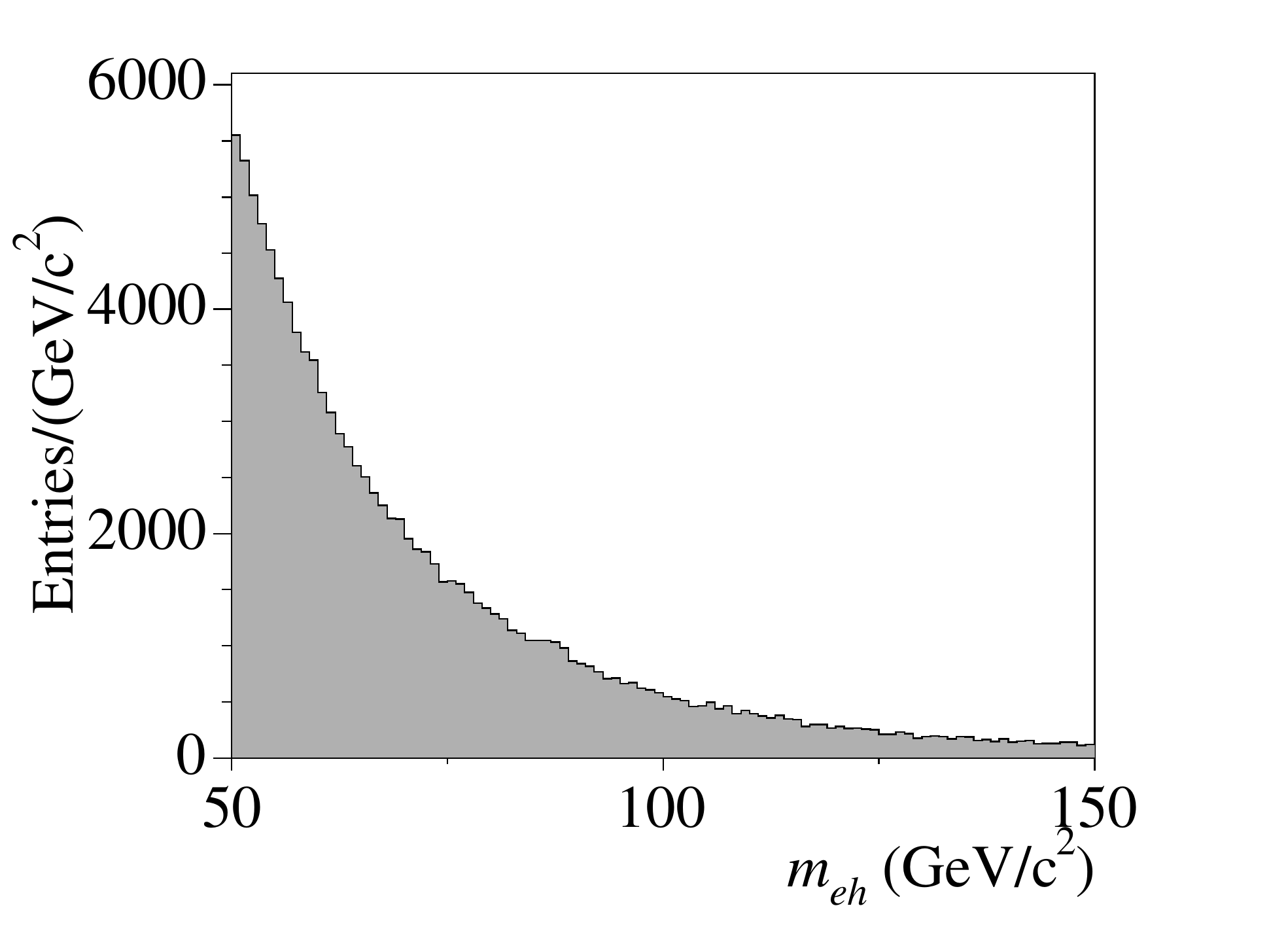}}
\caption{Invariant mass distribution of one central electron satisfying trigger
requirements and one hadronic candidate in data. \label{fig:trigfake}}
\end{figure}

\begin{figure*}[tb]
\centering
\mbox{
\mbox{\includegraphics[width=7.3cm,clip=]{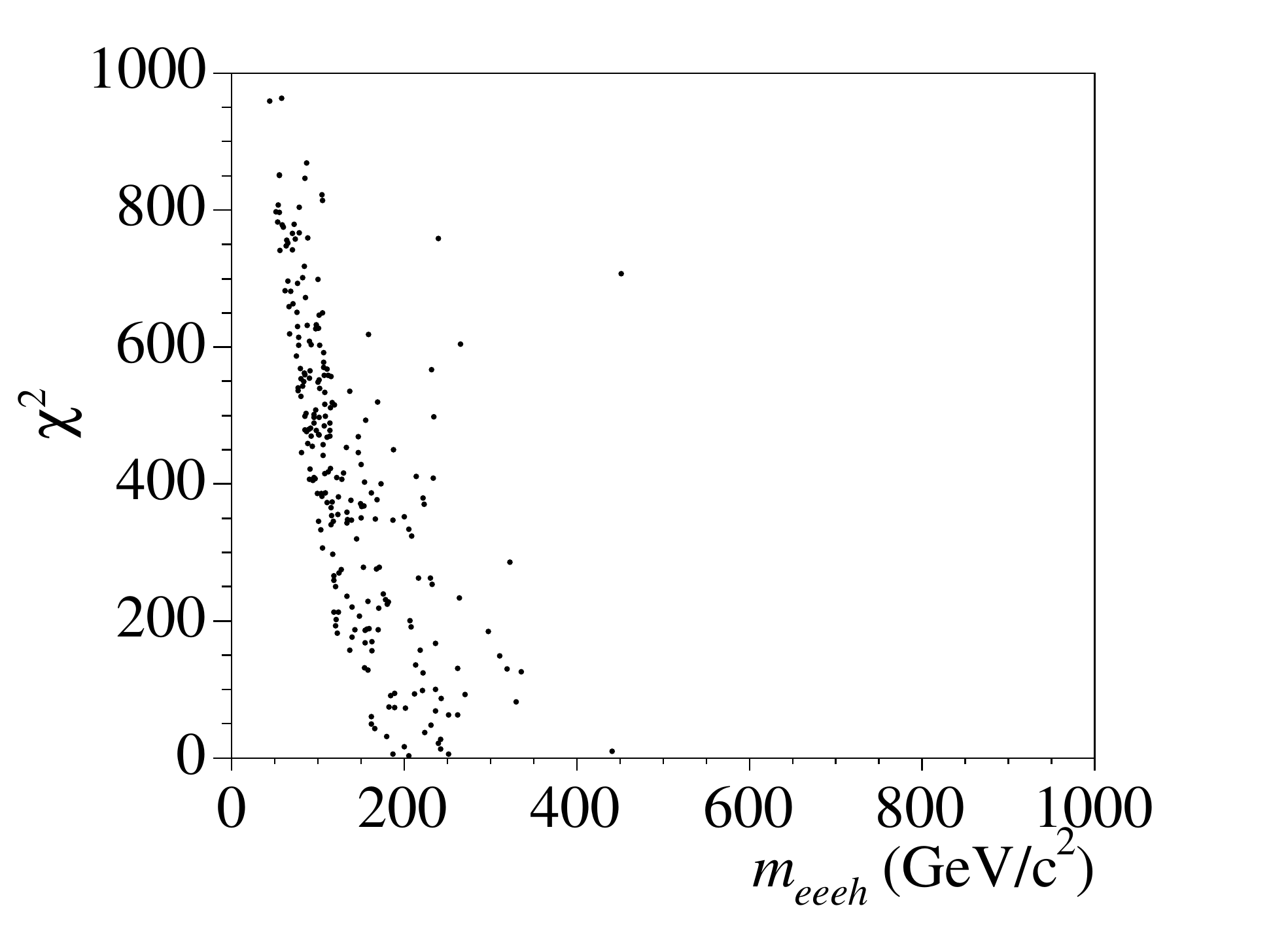}}
\mbox{\includegraphics[width=7.3cm,clip=]{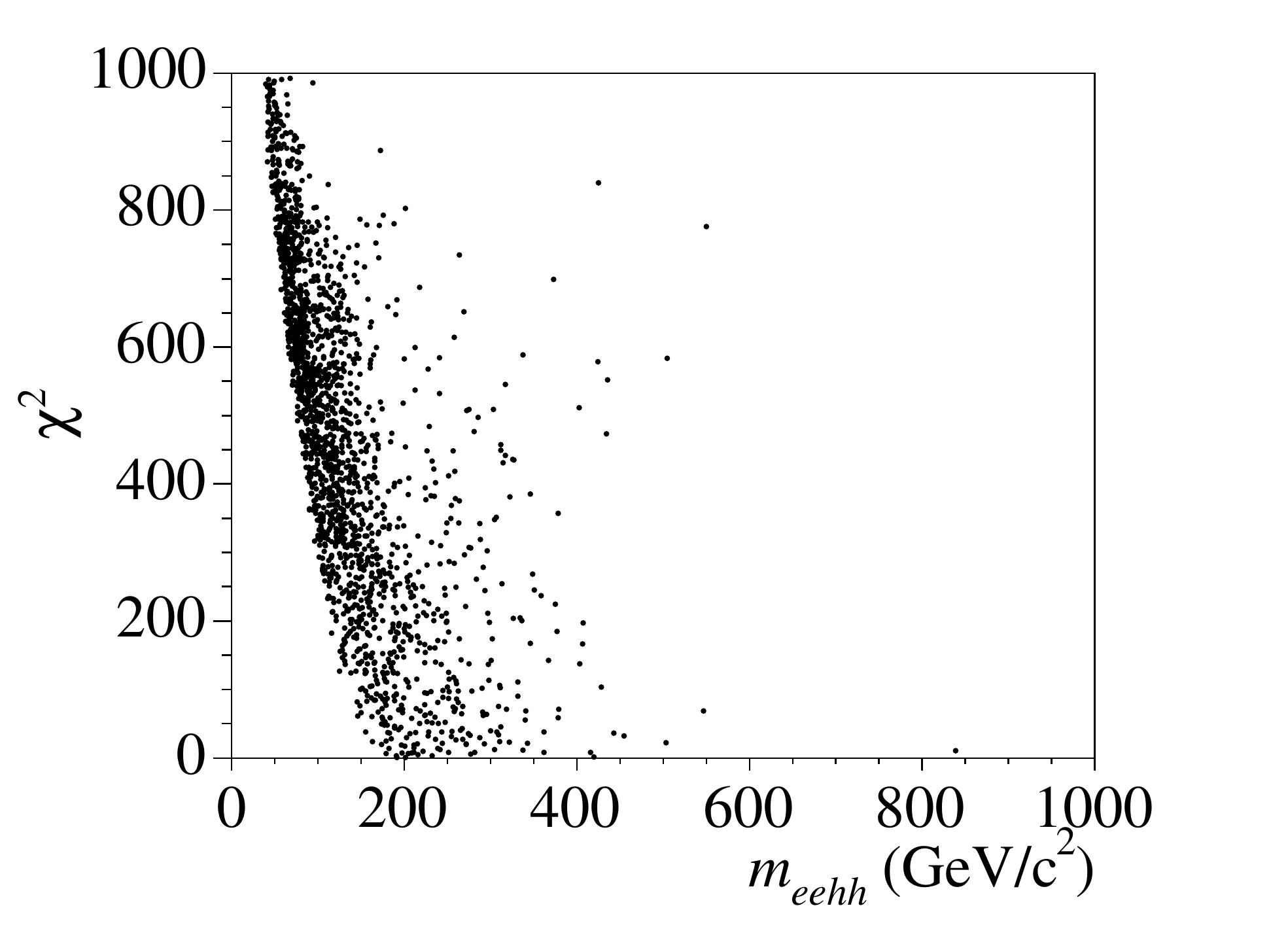}}
	}
\mbox{
	\mbox{\includegraphics[width=7.3cm,clip=]{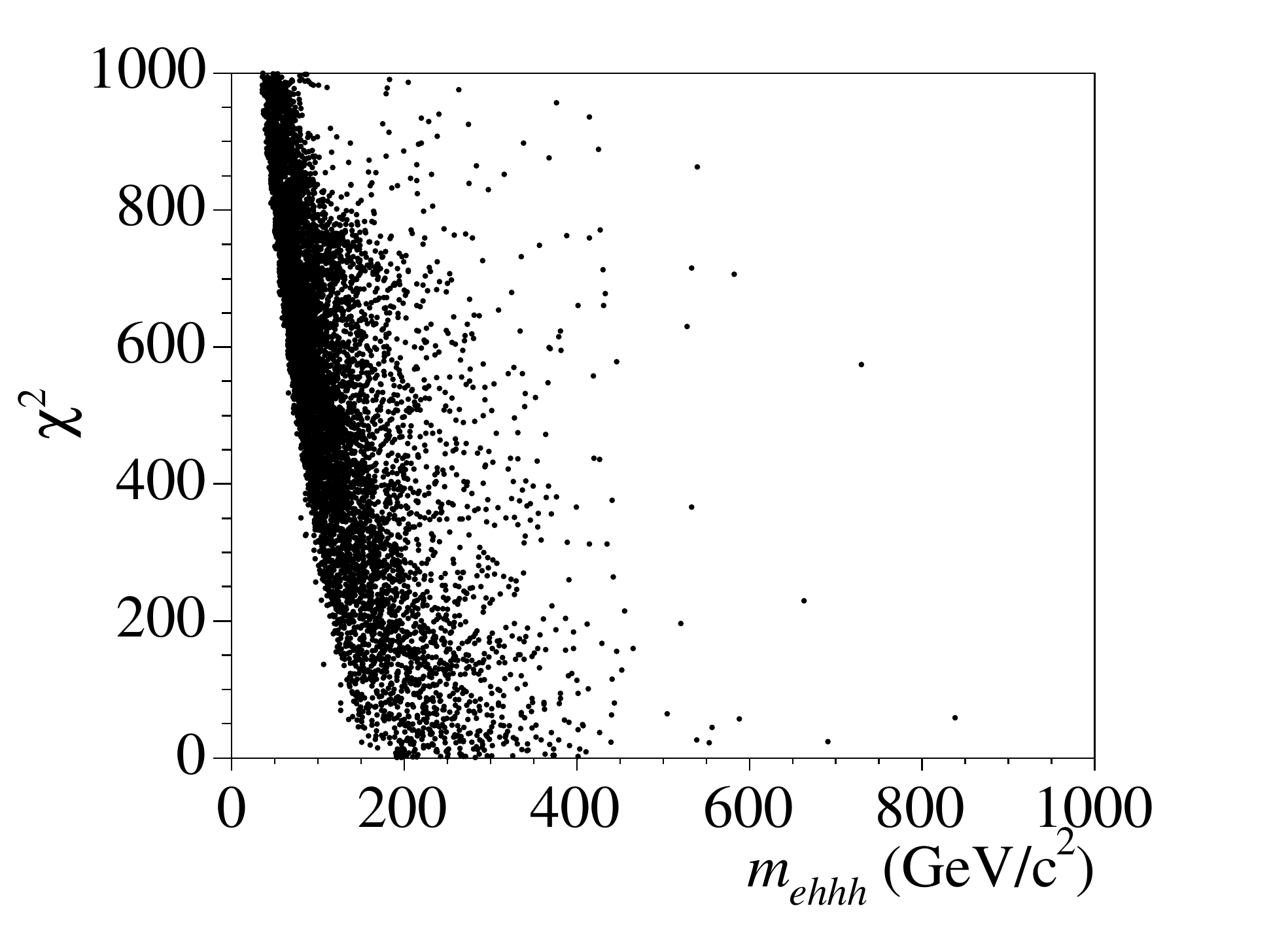}} 
	\mbox{\includegraphics[width=7.3cm,clip=]{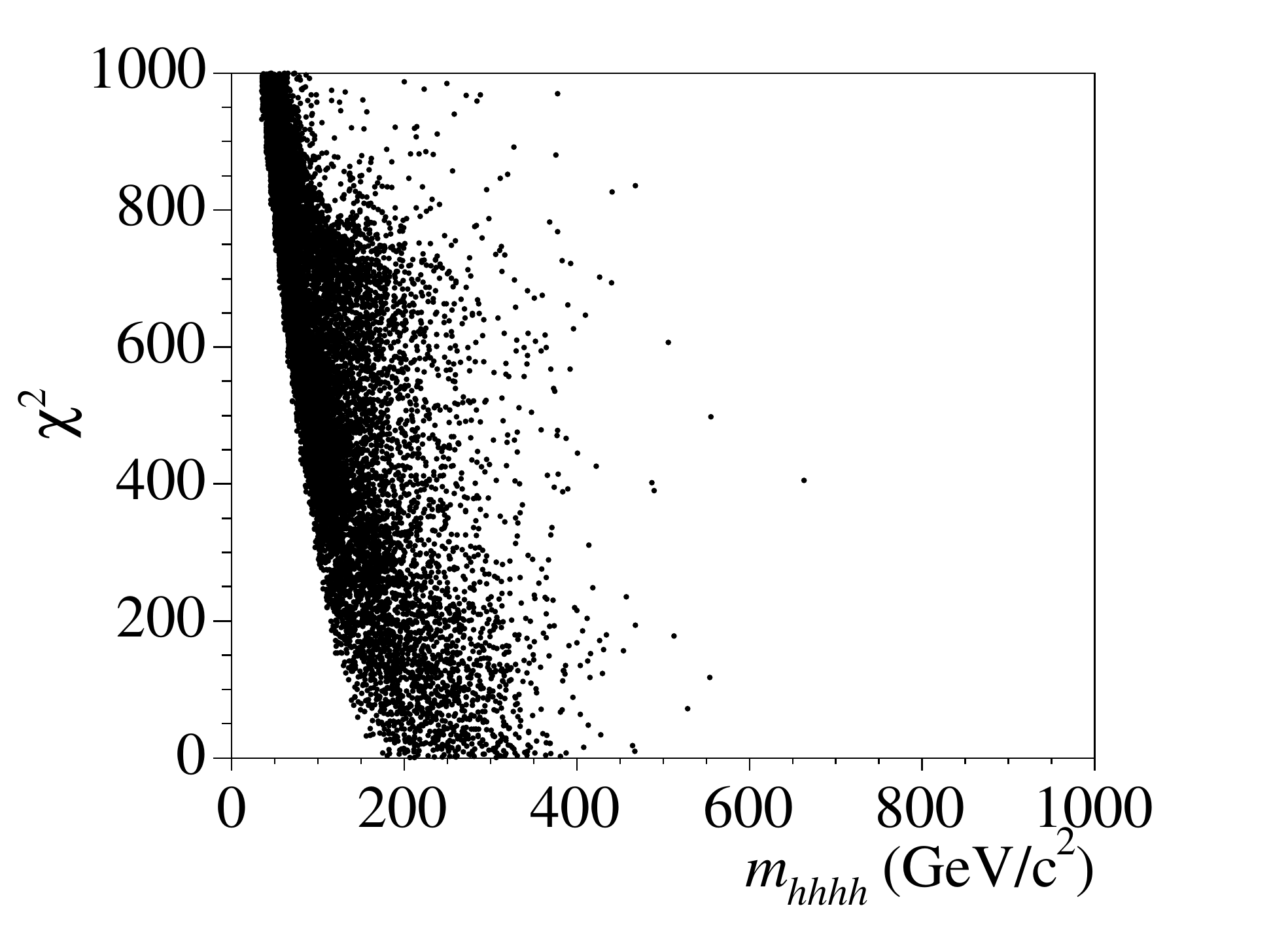}}
	}

\caption{
Distribution of \chisq vs \meeee for control samples containing one, two,
three, or four hadron candidates
with the number of events in the plot increasing with the number of
hadron candidates used in the combination.\label{fig:data_chisq_meeee} }
\end{figure*}

The single probability density function,
\begin{linenomath}
\begin{eqnarray}
f(\chisq,\meeee) & = & C \meeee^\gamma  e^{\chisq  \tau},
\end{eqnarray}
\end{linenomath}
where $C$ is a normalization constant,
provides an empirical description of the \chisq vs. \meeee distributions for
each of the four hadron-enriched control samples.  We obtain the parameters
$\gamma = -4.57 \pm 0.06$ and $\tau = -0.00319 \pm 0.00007$ from a
two-dimensional unbinned maximum likelihood fit to the low-mass four-electron
control region and the hadron-enriched control samples simultaneously, using
events with invariant mass above 185~\gevcc ($\sim 2 \times m_{Z^0}$.)  The control
samples containing mostly hadron candidates dominate the fit.
Fig.~\ref{fig:fit_meeee_fakes} shows the projections of the fit result in
the invariant mass dimension along with the data for each hadron-enriched sample.
The low-mass control region ($\meeee < 400$~\gevcc) contains five events
with $\meeee > 185$~\gevcc and serves to normalize the prediction of
background in the high-mass search region.  We integrate the fit result
above 500~\gevcc and $\chisq < 50$ to extract an estimate of the total
background for the high-mass region.  Using this method, we estimate $0.020
\pm 0.009$ (stat) $\pm~0.007$ (syst) background events from hadrons
misidentified as electrons in the search region.  The systematic uncertainty
on the background estimate is obtained by varying the functional form of the
probability density function fitted to the \meeee spectrum.

\begin{figure*}[tb]
\begin{center}
\mbox{
\mbox{\includegraphics[width=7.3cm,clip=]{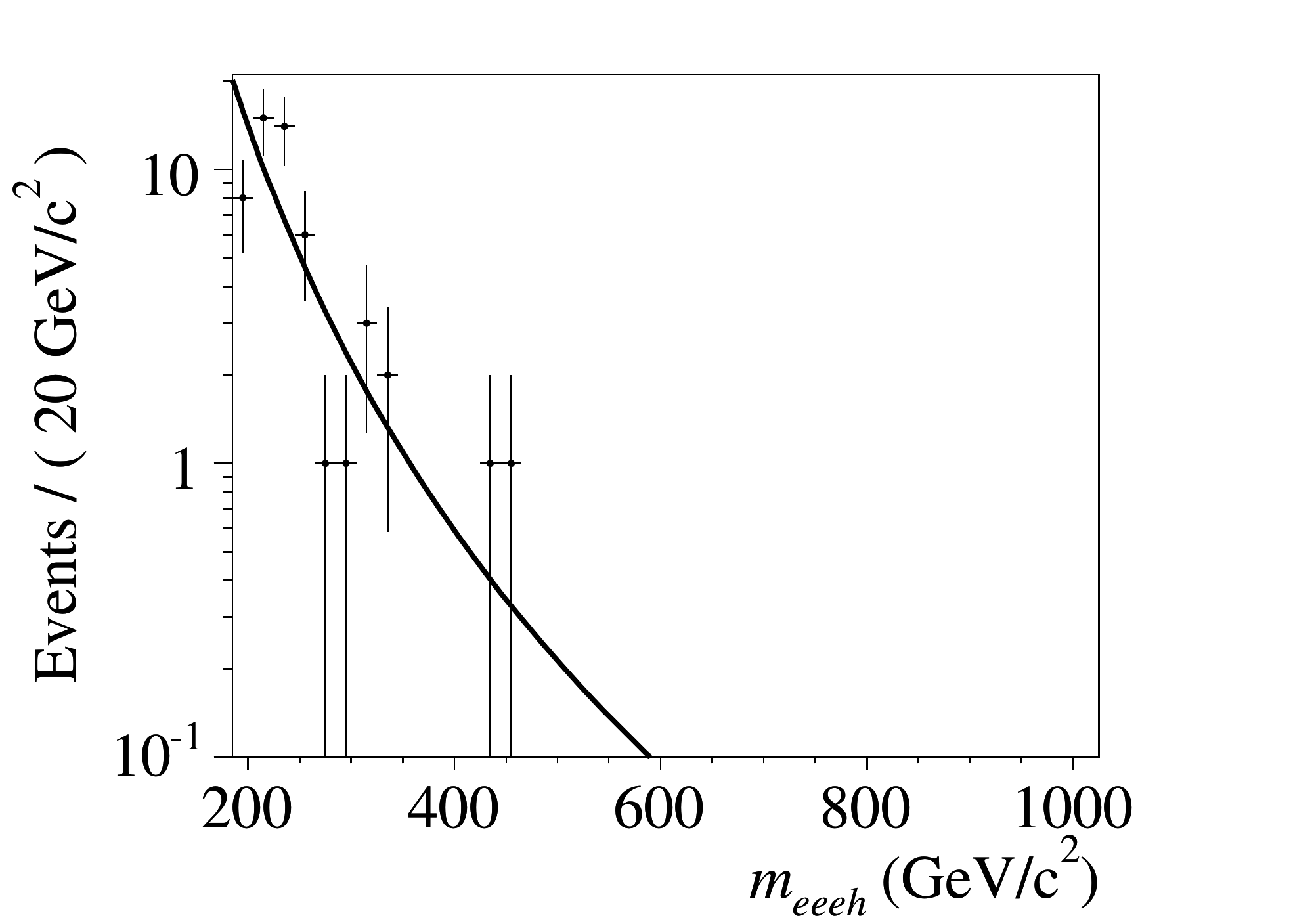} }
\mbox{\includegraphics[width=7.3cm,clip=]{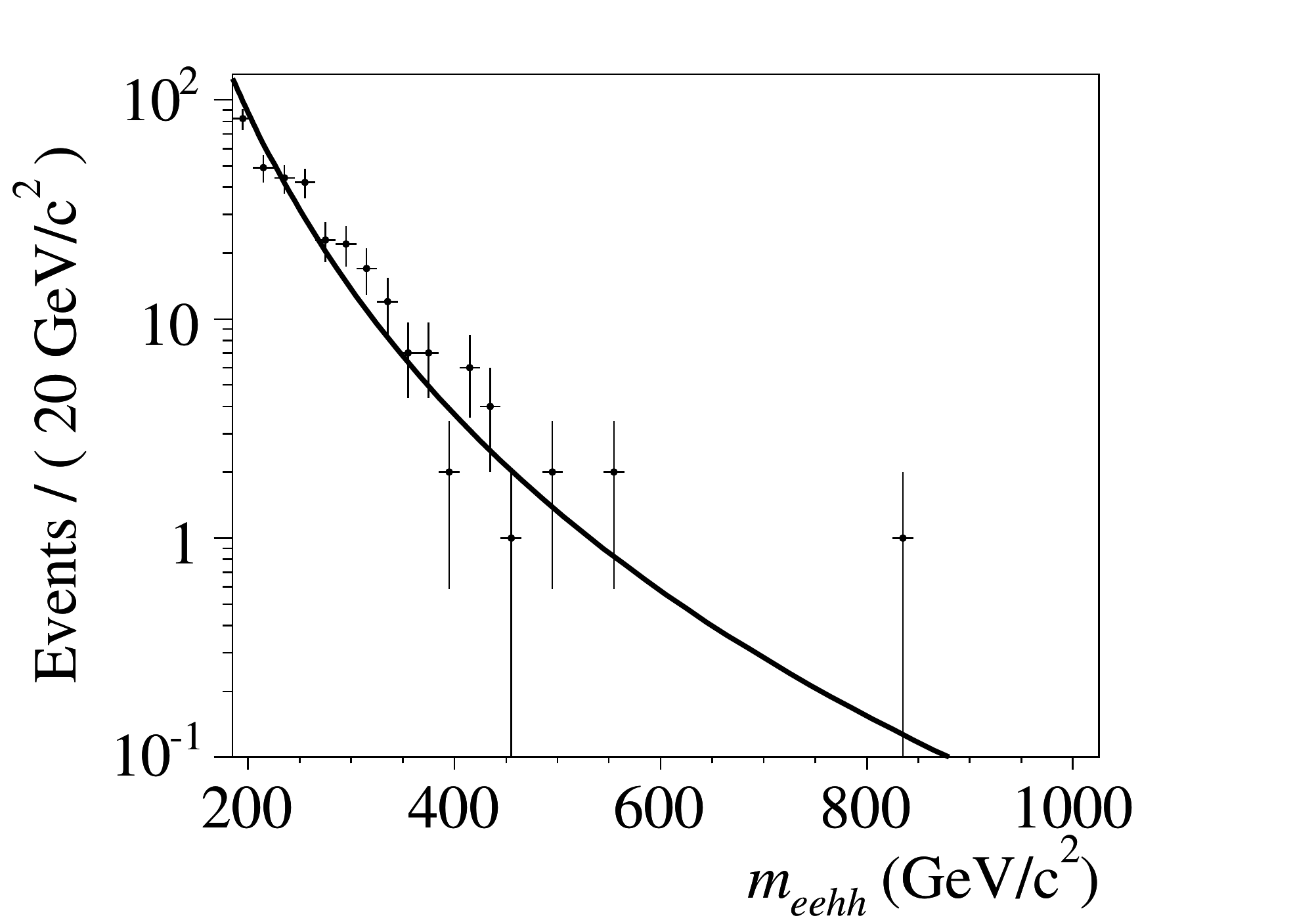} }
}
\mbox{
\mbox{ \includegraphics[width=7.3cm,clip=]{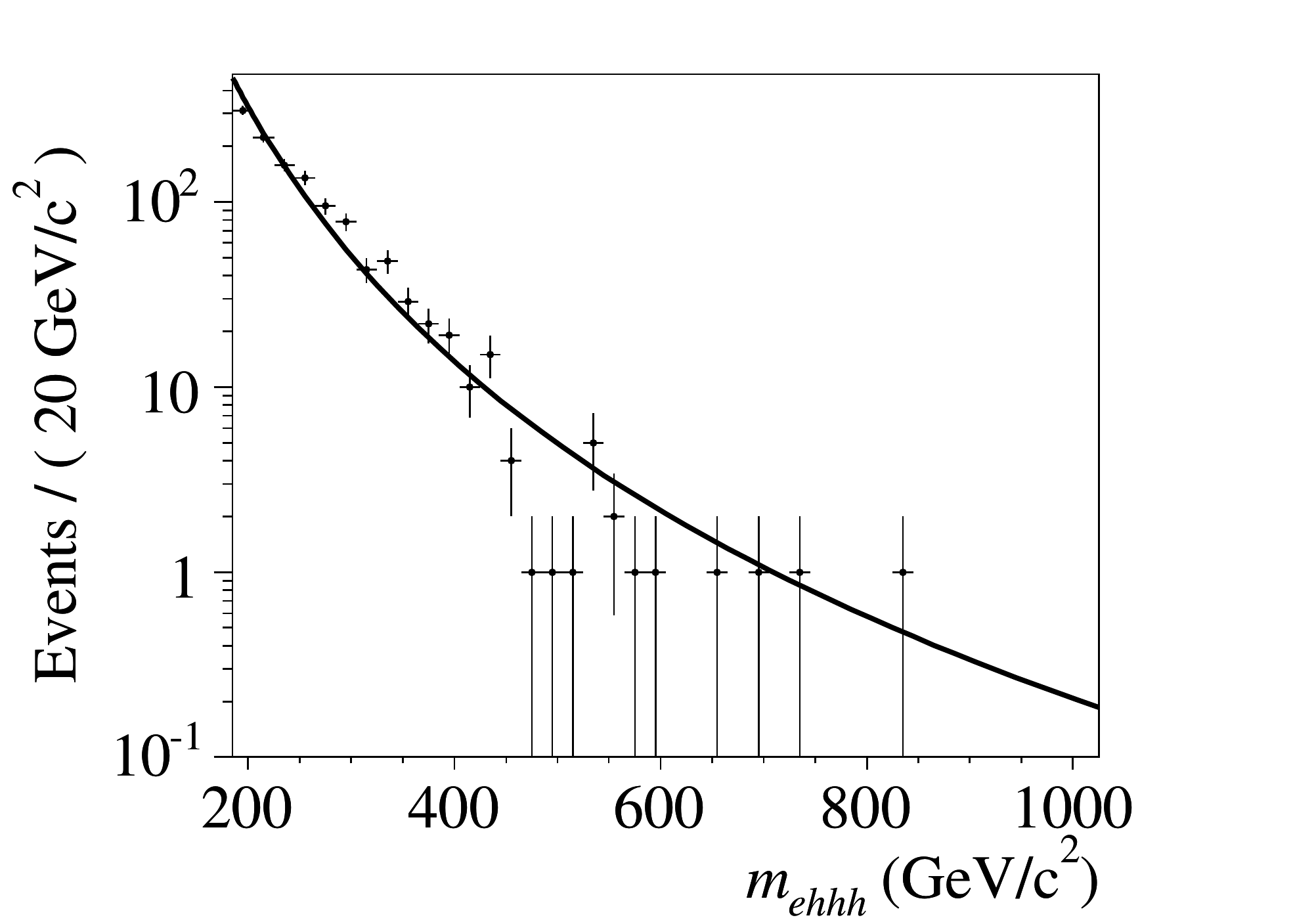} }
\mbox{ \includegraphics[width=7.3cm,clip=]{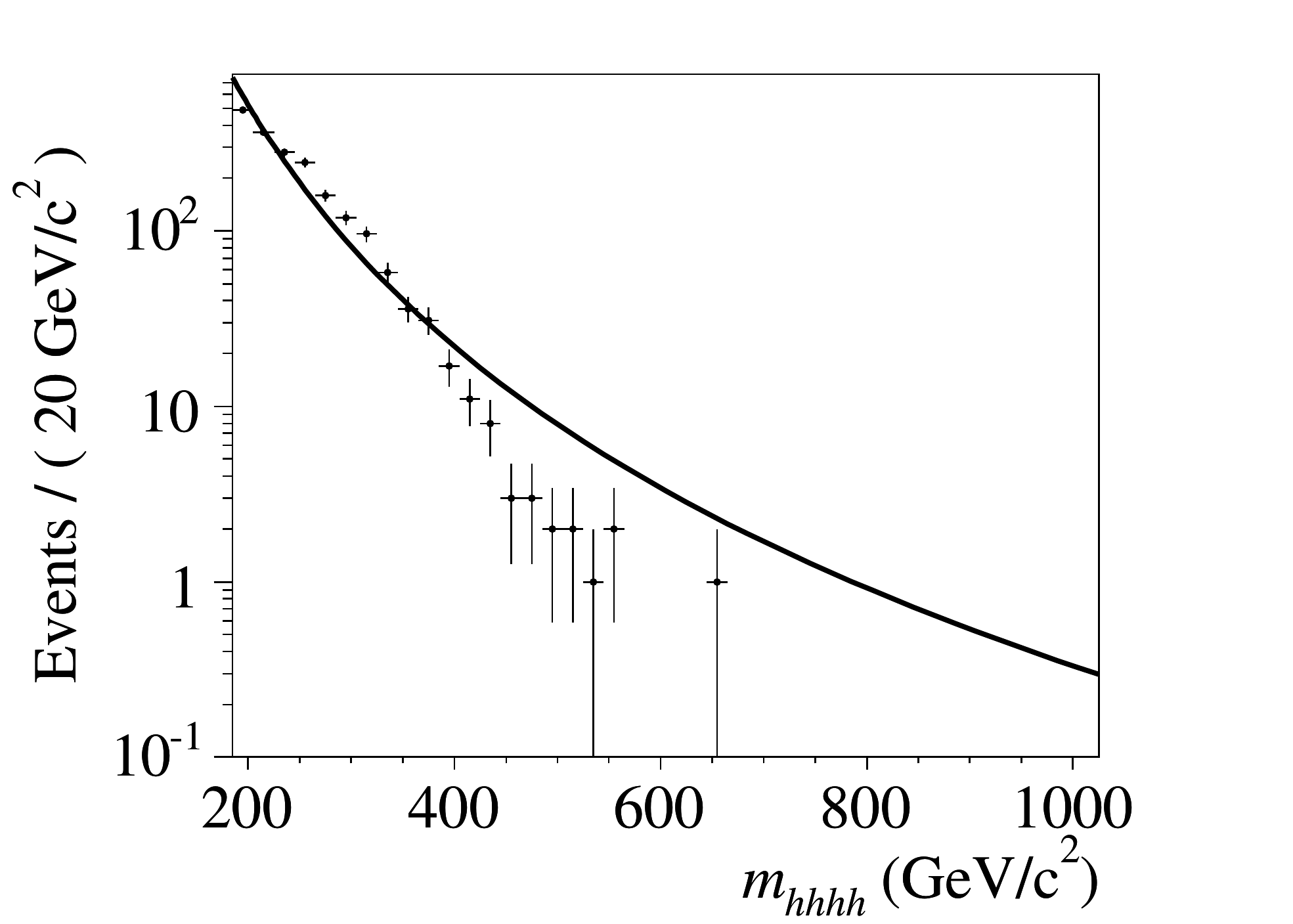} }
}
\end{center}
\caption{
Projections of fit to invariant mass in control samples of varying number of
electron and hadronic candidates with same ordering as in
Fig.~\ref{fig:data_chisq_meeee}.  Data are shown with the fit projection
overlaid.
\label{fig:fit_meeee_fakes}}
\end{figure*}

We have performed several cross checks to ensure that the fit provides a
reasonable estimate of the background rate at high mass. In particular, we
have performed the fit allowing the power-law parameter to vary independently for each of
the categories.  The parameters resulting from this fit along with the
number of events observed in each sample are shown in
Table~\ref{tab:fit:floatgamma}.  The fitted values for
$\gamma$ are consistent within errors across categories and with the nominal
result.  We have verified that the background shape in \meeee is independent
of \chisq in subsets of data in bins of
\chisq and have checked that the projected fit result is consistent with data.
\begin{table}
\caption[Fit result, gamma floating independently for each control sample. ]
{Result of fit with $\gamma$ floating independently for
each control sample.}
\label{tab:fit:floatgamma}
\begin{center}
\input{fit_floatgamma}
\end{center}
\end{table}

Standard model production of \zzerozzero~\cite{Campbell:1999ah} is the only
background to this search with two real \zzero bosons and possibly
four electrons in the final state.  While we
use data to estimate the total background from misidentified electrons in
the search region, we have studied the background from this source at
high mass with simulated events.  We determine geometric and
kinematic acceptance using Monte Carlo events generated by
\pythia~\cite{Sjostrand:2006za}, followed by a \geant-based simulation of
the \cdfii detector.  The expected number of events from this background
component is determined as the product of the cross section, the luminosity
of the sample, and the acceptance of the detector.  We estimate a total of
$0.54 \pm 0.04$ events in the four-electron sample.  In the invariant mass
region above 500~\gevcc, we expect $0.008 \pm 0.006$ events.  We estimate
the total background from production of standard model \zzerozzero events
and events in which hadrons are misidentified as electrons is $0.028 \pm
0.009$ (stat) $\pm~0.011$~(syst) events.

\section{Other Systematic Uncertainties}\label{sec:systematics}

When setting the cross-section limit, we have considered other systematic
uncertainties from several sources. The dominant source of these
uncertainties is from the measured luminosity (5.9\%)~\cite{lum}.
Other sources include
parton distribution function uncertainties (0.4\%), signal Monte Carlo
statistics (1.3\%), initial state radiation(1.0\%), and the
difference between electron identification efficiency in data and simulation
(1.0\% per electron).  The total systematic uncertainty from all these sources
is~7.3\%.

\section{Results \label{sec:results}}
The distribution of data events surviving all requirements is shown in
Fig.~\ref{fig:sig_chi2_meeee}.  We observe no events in the high-mass
signal region.  There is one event in the low-mass region with very small
\chisq
consistent with
standard model \zzerozzero production, while we expect $0.54$ standard model
\zzerozzero events over the entire mass range.  In the observed event, the total
invariant mass is 190~\gevcc, and the two \zzero candidates in the lowest
\chisq combination have measured masses of 91 and 92~\gevcc.
\begin{figure}[tb]
\centering
\includegraphics[width=7.3cm,clip=]{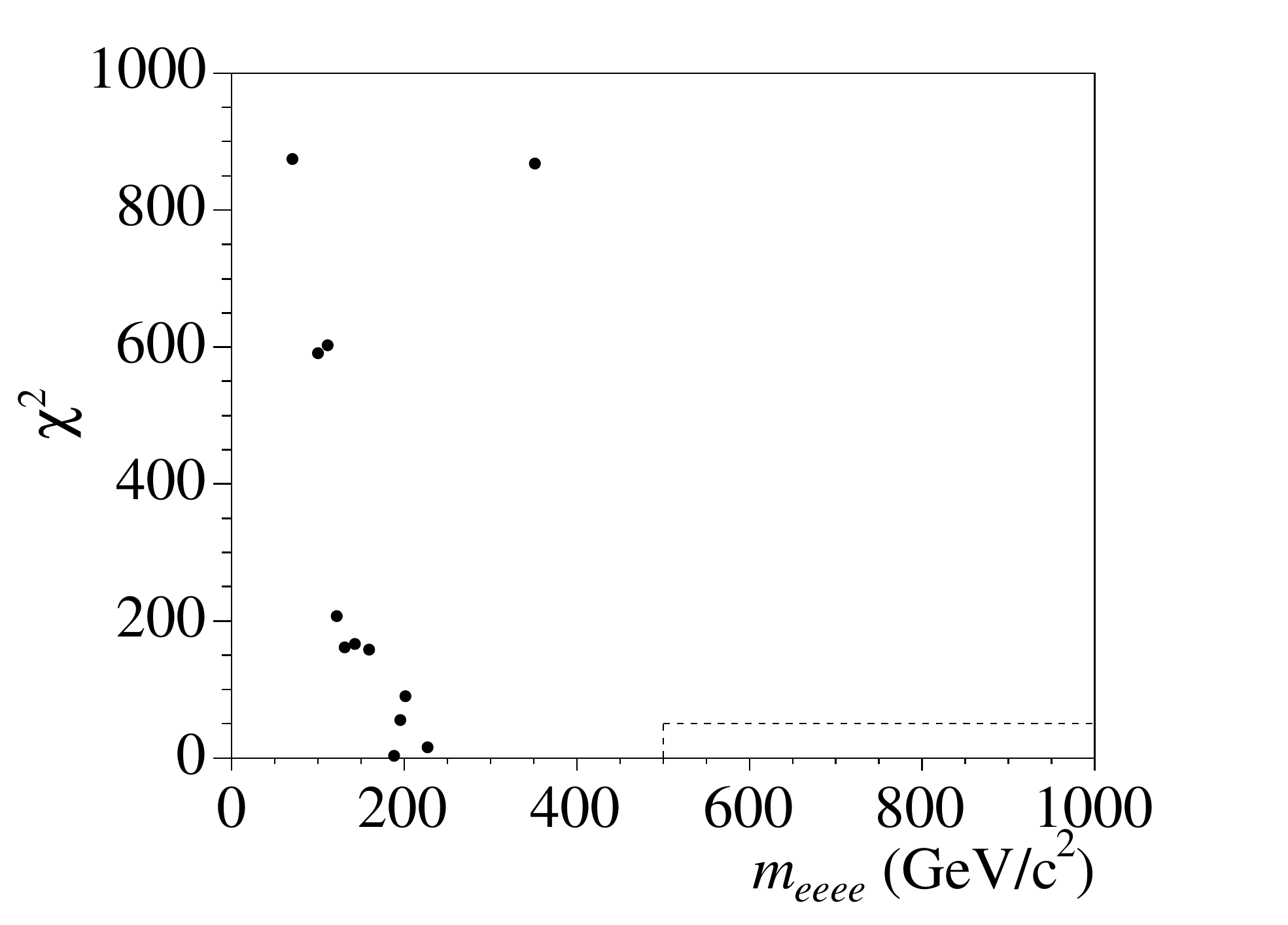}
\caption{
Distribution of \chisq vs \meeee for four-electron candidates in data.
The signal region is $\chisq<50$ and $\meeee > 500$~\gevcc, the boundaries
of which are shown by the dashed line.
\label{fig:sig_chi2_meeee}}
\end{figure}

We have set limits on $\sigma(\ppbar \rightarrow G) \times
\mathrm{BF}(G\rightarrow \zzero\zzero)$ in the context of a RS1 graviton scenario.
We use a Bayesian binned maximum likelihood method to extract 95\%
confidence level limits in 100~\gevcc wide windows centered on each mass.
We incorporate the effects of uncertainty on the background and signal
acceptance with a flat prior for the signal rate and Gaussian priors for the
acceptance and expected background.
The limits including systematic uncertainties on the acceptance
on $\sigma \times \mathrm{BF}(G\rightarrow \zzero \zzero)$ range from  $4
- 6$ pb, depending on the graviton mass, and   are shown in
Fig.~\ref{fig:limits} along with the prediction from the RS1 model for
\kovermp = 0.1.
In the future, the sensitivity of this search
will improve with more data as well as
additional acceptance from other \zzero decays.

\begin{figure}[tb]
\centering
\mbox{\includegraphics[width=7.3cm,clip=]{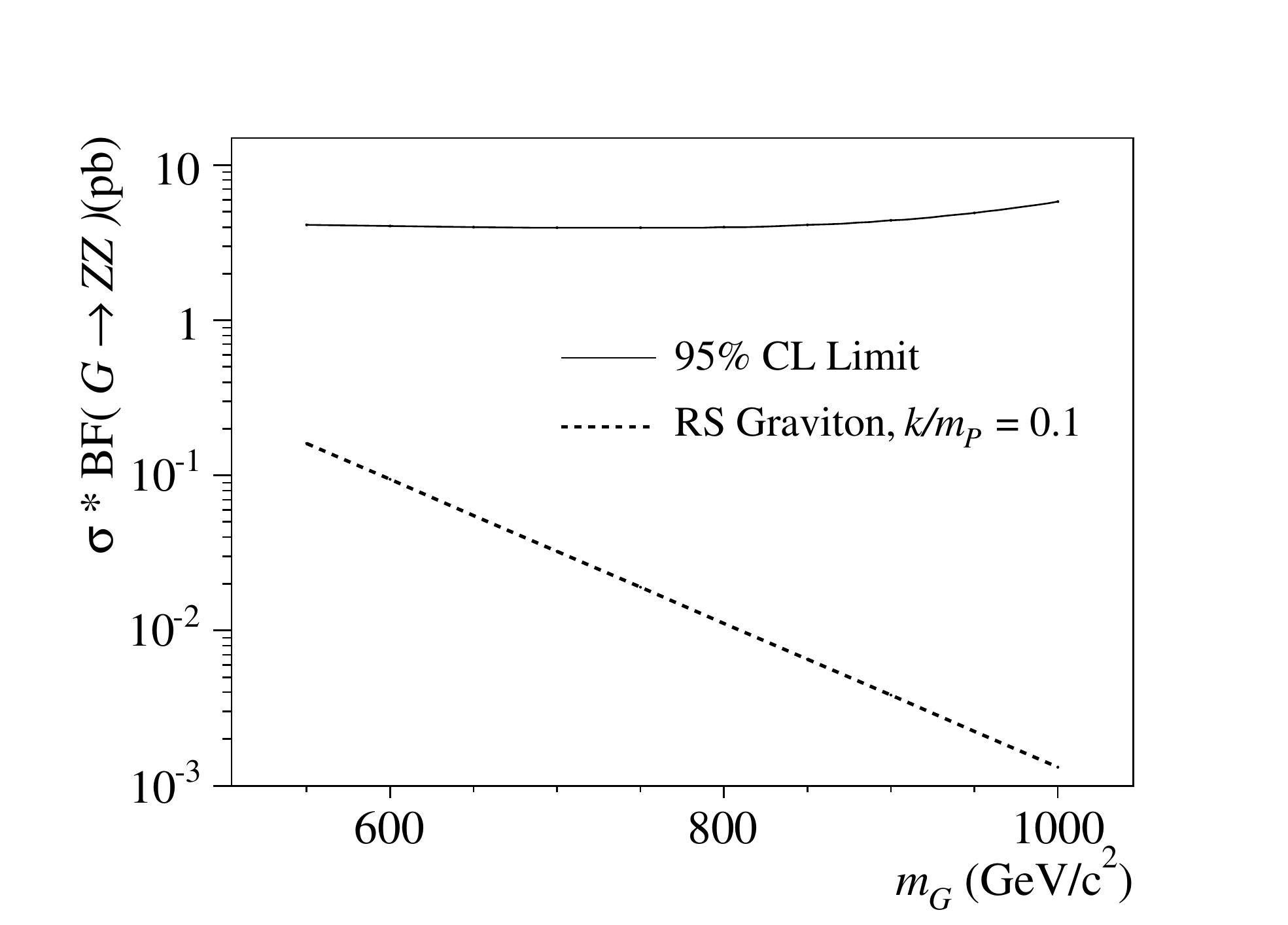}} 
  \caption{Limits on $\sigma \times \mathrm{BF}(G\rightarrow \zzero
  \zzero)$ versus mass.\label{fig:limits}}
\end{figure}

\section{Conclusions}
We have searched for production of particles decaying to a pair of \zzero
bosons.  We have estimated backgrounds from misidentified electrons using a
data-based technique, and the background from
standard model processes involving four electrons with simulations.
Using an optimized electron selection, we expect
$0.028 \pm 0.009$ (stat) $\pm 0.011$ (syst) total background events with
$\chisq < 50$ above 500~\gevcc in 1.10~\ifb, and observe no events.  In the
absence of evidence for a signal, we have set limits on $\sigma({p \bar{p}
\rightarrow G (1.96 \gev)})
\times \mathrm{BF}(G\rightarrow \zzerozzero)$. 

\ifprd
\begin{acknowledgments}
\else
\section{Acknowledgments}
\fi
We thank the Fermilab staff and the technical staffs of the participating
institutions for their vital contributions. This work was supported by the
U.S. Department of Energy and National Science Foundation; the Italian
Istituto Nazionale di Fisica Nucleare; the Ministry of Education, Culture,
Sports, Science and Technology of Japan; the Natural Sciences and
Engineering Research Council of Canada; the National Science Council of the
Republic of China; the Swiss National Science Foundation; the A.P. Sloan
Foundation; the Bundesministerium f\"ur Bildung und Forschung, Germany; the
Korean Science and Engineering Foundation and the Korean Research
Foundation; the Science and Technology Facilities Council and the Royal
Society, UK; the Institut National de Physique Nucleaire et Physique des
Particules/CNRS; the Russian Foundation for Basic Research; the Comisi\'on
Interministerial de Ciencia y Tecnolog\'{\i}a, Spain; the European
Community's Human Potential Programme; the Slovak R\&D Agency; and the
Academy of Finland.
\ifprd
\end{acknowledgments}
\fi

\end{document}

%% file: July_2007_Authors_Visitors.tex
\affiliation{Institute of Physics, Academia Sinica, Taipei, Taiwan 11529, Republic of China} 
\affiliation{Argonne National Laboratory, Argonne, Illinois 60439} 
\affiliation{Institut de Fisica d'Altes Energies, Universitat Autonoma de Barcelona, E-08193, Bellaterra (Barcelona), Spain} 
\affiliation{Baylor University, Waco, Texas  76798} 
\affiliation{Istituto Nazionale di Fisica Nucleare, University of Bologna, I-40127 Bologna, Italy} 
\affiliation{Brandeis University, Waltham, Massachusetts 02254} 
\affiliation{University of California, Davis, Davis, California  95616} 
\affiliation{University of California, Los Angeles, Los Angeles, California  90024} 
\affiliation{University of California, San Diego, La Jolla, California  92093} 
\affiliation{University of California, Santa Barbara, Santa Barbara, California 93106} 
\affiliation{Instituto de Fisica de Cantabria, CSIC-University of Cantabria, 39005 Santander, Spain} 
\affiliation{Carnegie Mellon University, Pittsburgh, PA  15213} 
\affiliation{Enrico Fermi Institute, University of Chicago, Chicago, Illinois 60637} 
\affiliation{Comenius University, 842 48 Bratislava, Slovakia; Institute of Experimental Physics, 040 01 Kosice, Slovakia} 
\affiliation{Joint Institute for Nuclear Research, RU-141980 Dubna, Russia} 
\affiliation{Duke University, Durham, North Carolina  27708} 
\affiliation{Fermi National Accelerator Laboratory, Batavia, Illinois 60510} 
\affiliation{University of Florida, Gainesville, Florida  32611} 
\affiliation{Laboratori Nazionali di Frascati, Istituto Nazionale di Fisica Nucleare, I-00044 Frascati, Italy} 
\affiliation{University of Geneva, CH-1211 Geneva 4, Switzerland} 
\affiliation{Glasgow University, Glasgow G12 8QQ, United Kingdom} 
\affiliation{Harvard University, Cambridge, Massachusetts 02138} 
\affiliation{Division of High Energy Physics, Department of Physics, University of Helsinki and Helsinki Institute of Physics, FIN-00014, Helsinki, Finland} 
\affiliation{University of Illinois, Urbana, Illinois 61801} 
\affiliation{The Johns Hopkins University, Baltimore, Maryland 21218} 
\affiliation{Institut f\"{u}r Experimentelle Kernphysik, Universit\"{a}t Karlsruhe, 76128 Karlsruhe, Germany} 
\affiliation{Center for High Energy Physics: Kyungpook National University, Daegu 702-701, Korea; Seoul National University, Seoul 151-742, Korea; Sungkyunkwan University, Suwon 440-746, Korea; Korea Institute of Science and Technology Information, Daejeon, 305-806, Korea; Chonnam National University, Gwangju, 500-757, Korea} 
\affiliation{Ernest Orlando Lawrence Berkeley National Laboratory, Berkeley, California 94720} 
\affiliation{University of Liverpool, Liverpool L69 7ZE, United Kingdom} 
\affiliation{University College London, London WC1E 6BT, United Kingdom} 
\affiliation{Centro de Investigaciones Energeticas Medioambientales y Tecnologicas, E-28040 Madrid, Spain} 
\affiliation{Massachusetts Institute of Technology, Cambridge, Massachusetts  02139} 
\affiliation{Institute of Particle Physics: McGill University, Montr\'{e}al, Canada H3A~2T8; and University of Toronto, Toronto, Canada M5S~1A7} 
\affiliation{University of Michigan, Ann Arbor, Michigan 48109} 
\affiliation{Michigan State University, East Lansing, Michigan  48824} 
\affiliation{University of New Mexico, Albuquerque, New Mexico 87131} 
\affiliation{Northwestern University, Evanston, Illinois  60208} 
\affiliation{The Ohio State University, Columbus, Ohio  43210} 
\affiliation{Okayama University, Okayama 700-8530, Japan} 
\affiliation{Osaka City University, Osaka 588, Japan} 
\affiliation{University of Oxford, Oxford OX1 3RH, United Kingdom} 
\affiliation{University of Padova, Istituto Nazionale di Fisica Nucleare, Sezione di Padova-Trento, I-35131 Padova, Italy} 
\affiliation{LPNHE, Universite Pierre et Marie Curie/IN2P3-CNRS, UMR7585, Paris, F-75252 France} 
\affiliation{University of Pennsylvania, Philadelphia, Pennsylvania 19104} 
\affiliation{Istituto Nazionale di Fisica Nucleare Pisa, Universities of Pisa, Siena and Scuola Normale Superiore, I-56127 Pisa, Italy} 
\affiliation{University of Pittsburgh, Pittsburgh, Pennsylvania 15260} 
\affiliation{Purdue University, West Lafayette, Indiana 47907} 
\affiliation{University of Rochester, Rochester, New York 14627} 
\affiliation{The Rockefeller University, New York, New York 10021} 
\affiliation{Istituto Nazionale di Fisica Nucleare, Sezione di Roma 1, University of Rome ``La Sapienza," I-00185 Roma, Italy} 
\affiliation{Rutgers University, Piscataway, New Jersey 08855} 
\affiliation{Texas A\&M University, College Station, Texas 77843} 
\affiliation{Istituto Nazionale di Fisica Nucleare, University of Trieste/\ Udine, Italy} 
\affiliation{University of Tsukuba, Tsukuba, Ibaraki 305, Japan} 
\affiliation{Tufts University, Medford, Massachusetts 02155} 
\affiliation{Waseda University, Tokyo 169, Japan} 
\affiliation{Wayne State University, Detroit, Michigan  48201} 
\affiliation{University of Wisconsin, Madison, Wisconsin 53706} 
\affiliation{Yale University, New Haven, Connecticut 06520} 
\author{T.~Aaltonen}
\affiliation{Division of High Energy Physics, Department of Physics, University of Helsinki and Helsinki Institute of Physics, FIN-00014, Helsinki, Finland}
\author{J.~Adelman}
\affiliation{Enrico Fermi Institute, University of Chicago, Chicago, Illinois 60637}
\author{T.~Akimoto}
\affiliation{University of Tsukuba, Tsukuba, Ibaraki 305, Japan}
\author{M.G.~Albrow}
\affiliation{Fermi National Accelerator Laboratory, Batavia, Illinois 60510}
\author{B.~\'{A}lvarez~Gonz\'{a}lez}
\affiliation{Instituto de Fisica de Cantabria, CSIC-University of Cantabria, 39005 Santander, Spain}
\author{S.~Amerio}
\affiliation{University of Padova, Istituto Nazionale di Fisica Nucleare, Sezione di Padova-Trento, I-35131 Padova, Italy}
\author{D.~Amidei}
\affiliation{University of Michigan, Ann Arbor, Michigan 48109}
\author{A.~Anastassov}
\affiliation{Rutgers University, Piscataway, New Jersey 08855}
\author{A.~Annovi}
\affiliation{Laboratori Nazionali di Frascati, Istituto Nazionale di Fisica Nucleare, I-00044 Frascati, Italy}
\author{J.~Antos}
\affiliation{Comenius University, 842 48 Bratislava, Slovakia; Institute of Experimental Physics, 040 01 Kosice, Slovakia}
% ...
\collaboration{CDF Collaboration\footnote{With visitors from $^a$University of Athens, 15784 Athens, Greece, 
$^b$Chinese Academy of Sciences, Beijing 100864, China, 
$^c$University of Bristol, Bristol BS8 1TL, United Kingdom, 
$^d$University Libre de Bruxelles, B-1050 Brussels, Belgium, 
$^e$University of California Irvine, Irvine, CA  92697, 
$^f$University of California Santa Cruz, Santa Cruz, CA  95064, 
$^g$Cornell University, Ithaca, NY  14853, 
$^h$University of Cyprus, Nicosia CY-1678, Cyprus, 
$^i$University College Dublin, Dublin 4, Ireland, 
$^j$University of Edinburgh, Edinburgh EH9 3JZ, United Kingdom, 
$^k$University of Heidelberg, D-69120 Heidelberg, Germany, 
$^l$Universidad Iberoamericana, Mexico D.F., Mexico, 
$^m$University of Manchester, Manchester M13 9PL, England, 
$^n$Nagasaki Institute of Applied Science, Nagasaki, Japan, 
$^o$University de Oviedo, E-33007 Oviedo, Spain, 
$^p$Queen Mary, University of London, London, E1 4NS, England, 
$^q$Texas Tech University, Lubbock, TX  79409, 
$^r$IFIC(CSIC-Universitat de Valencia), 46071 Valencia, Spain, 
}}
\noaffiliation

%% file: fit_floatgamma.tex
\mbox{
\begin{tabular}{|c|c|c|}
\hline
Sample & Events & $\gamma$   \\
\hline \hline
$eeee$ & 5 & $-5.90 \pm 2.14 $\\
$eeeh$ & 52 & $-4.85 \pm 0.55 $\\
$eehh$ & 323 & $-4.28 \pm 0.19 $\\
$ehhh$ & 1208 & $-4.43 \pm 0.10 $\\
$hhhh$ & 1927 & $-4.71 \pm 0.09 $\\
\hline
\end{tabular}
}

%% file: prd.bbl
\begin{thebibliography}{99}
\bibitem{Abulencia:2007tu}
  A.~Abulencia {\it et al.}  (CDF Collaboration),
  Phys.\ Rev.\ Lett.\  {\bf 98}, 161801 (2007).

\bibitem{Abazov:2005ys}
  V.~M.~Abazov {\it et al.}  (D0 Collaboration),
  Phys.\ Rev.\ Lett.\  {\bf 95}, 141802 (2005).


\bibitem{Affolder:2001cd}
  A.~A.~Affolder {\it et al.}  (CDF Collaboration),
  Phys.\ Rev.\ Lett.\  {\bf 88}, 071806 (2002).


\bibitem{Acosta:2005pq}
  D.~E.~Acosta {\it et al.}  [CDF Collaboration],
  Phys.\ Rev.\  D {\bf 71}, 091105 (2005)
  [arXiv:hep-ex/0501021].

\bibitem{Randall:1999ee}
  L.~Randall and R.~Sundrum,
  Phys.\ Rev.\ Lett.\  {\bf 83}, 3370 (1999).

\bibitem{Randall:1999vf}
L.~Randall and R.~Sundrum,
Phys.\ Rev.\ Lett.\  {\bf 83}, 4690 (1999).

\bibitem{Allanach:2002gn}
  B.~C.~Allanach, K.~Odagiri, M.~J.~Palmer, M.~A.~Parker, A.~Sabetfakhri and B.~R.~Webber,
  JHEP {\bf 0212}, 039 (2002)

\bibitem{:2007kpa}
  T.~Aaltonen {\it et al.}  [CDF Collaboration],
  Phys.\ Rev.\ Lett.\  {\bf 99}, 171801 (2007)


\bibitem{:2007sb}
  T.~Aaltonen {\it et al.}  (CDF Collaboration),
  arXiv:hep-ex/07072524.

\bibitem{Abazov:2005pi}
  V.~M.~Abazov {\it et al.}  (D0 Collaboration),
  Phys.\ Rev.\ Lett.\  {\bf 95}, 091801 (2005).

\bibitem{Fitzpatrick:2007qr}
  A.~L.~Fitzpatrick, J.~Kaplan, L.~Randall, and L.~T.~Wang,
  J. High Energy Phys. {\bf 0709}, 013 (2007).

%
%

\bibitem{Acosta:2004yw}
  D.~Acosta {\it et al.}  (CDF Collaboration),
  Phys.\ Rev.\  D {\bf 71}, 032001 (2005).
%
\bibitem{Sill:2000zz}
  A.~Sill {\it et al.}, 
  Nucl.\ Instrum.\ Methods\  A {\bf 447}, 1 (2000).
  A.~A.~Affolder {\it et al.}, 
  Nucl.\ Instrum.\ Methods\  A {\bf 453}, 84 (2000).

\bibitem{coordinates}
CDF uses a cylindrical coordinate system in which $\theta$ ($\phi$) is the
polar (azimuthal) angle, $r$ is the radius from the nominal beam axis, and
$+z$ points along the proton beam direction and is zero at the center of the
detector.  The pseudorapidity is defined as $\eta \equiv -\ln
\tan(\theta/2)$.  Energy (momentum) transverse to the beam is defined $\Et \equiv E \sin
\theta$ ($\pt \equiv p \sin \theta$), where $E$ is energy measured by the
calorimeter and $p$ is momentum measured by the spectrometer.

\bibitem{Affolder:2003ep}
  A.~A.~Affolder {\it et al.}, 
  Nucl.\ Instrum.\ Methods\  A {\bf 526}, 249 (2004).

\bibitem{Balka:1987ty}
  L.~Balka {\it et al.}, 
  Nucl.\ Instrum.\ Methods\  A {\bf 267}, 272 (1988).

\bibitem{clc}
  D.~Acosta {\it et al.},
  Nucl.\ Instrum.\ Methods\  A{\bf 494}, 57 (2002).


\bibitem{Yao:2006px}
  W.-M.~Yao {\it et al.},
  J.\ Phys.\ G {\bf 33}, 1 (2006).

\bibitem{Brun:1987ma} 
  R.~Brun, F.~Bruyant, M.~Maire, A.~C.~McPherson and P.~Zanarini,
  ``GEANT3,''
CERN Report No. CERN-DD/EE/84-1 (unpublished).

\bibitem{Corcella:2000bw}
  G.~Corcella {\it et al.},
  ``HERWIG 6: An event generator for hadron emission reactions with
  interfering gluons (including supersymmetric processes),''
  J. High Energy Phys {\bf 0101}, 010 (2001).

%
\bibitem{Campbell:1999ah}
 J.~M.~Campbell and R.~K.~Ellis,
 Phys.\ Rev.\  D {\bf 60}, 113006 (1999).

\bibitem{Sjostrand:2006za}
  T.~Sjostrand, S.~Mrenna, and P.~Skands,
``PYTHIA 6.4 physics and manual,''
  J. High Energy Phys. {\bf 0605}, 026 (2006).

\bibitem{lum}
S. Klimenko, J. Konigsberg, and T.M. Liss, FERMILAB-FN-0741 (2003) (unpublished).

\end{thebibliography}
